# The INTERSPEECH 2020 Deep Noise Suppression Challenge: Datasets, Subjective Speech Quality and Testing Framework

*Chandan K. A. Reddy[1], Ebrahim Beyrami[1], Harishchandra Dubey[1], Vishak Gopal[1], Roger Cheng[1], Ross Cutler[1], Sergiy Matusevych[1], Robert Aichner[1], Ashkan Aazami[1], Sebastian Braun[2], Puneet Rana[1], Sriram Srinivasan[1], Johannes Gehrke[1]*

[1]Microsoft Corporation, Redmond, WA, USA
[2]Microsoft Research, Redmond, WA, USA

`{Chandan.karadagur, firstname.lastname}@microsoft.com`

## Abstract

The INTERSPEECH 2020 Deep Noise Suppression Challenge is intended to promote collaborative research in real-time single-channel Speech Enhancement aimed to maximize the subjective (perceptual) quality of the enhanced speech. A typical approach to evaluate the noise suppression methods is to use objective metrics on the test set obtained by splitting the original dataset. Many publications report reasonable performance on the synthetic test set drawn from the same distribution as that of the training set. However, often the model performance degrades significantly on real recordings. Also, most of the conventional objective metrics do not correlate well with subjective tests and lab subjective tests are not scalable for a large test set. In this challenge, we open source a large clean speech and noise corpus for training the noise suppression models and a representative test set to real-world scenarios consisting of both synthetic and real recordings. We also open source an online subjective test framework based on ITU-T P.808 for researchers to quickly test their developments. The winners of this challenge will be selected based on subjective evaluation on a representative test set using P.808 framework.

## 1. Introduction

As the number of people working remotely and in open office environments continue to increase, the desire to have a video/audio call with excellent speech quality and intelligibility has become more important than ever before. The degradation of speech quality due to background noise is one of the major sources for poor quality ratings in voice calls. The conventional Speech Enhancement (SE) techniques are based on statistical models estimated from the noisy observations. These methods perform well on stationary noises but fail to effectively suppress non-stationary noises [1]–[5]. Recently, SE is treated as a supervised learning problem in which the patterns within speech and noise are learned using the training data [6]. Deep Neural Networks (DNN) are used to estimate speech in either spectral or time domain. DNN based methods are shown to outperform conventional SE techniques in suppressing non-stationary noises [7]–[10].

Over four decades of research in noise suppression techniques led to development of signal processing and deep neural networks-based approaches that use time-domain audio signal or spectral features [1]–[5] for estimating the multiplicative mask for noise reduction. Most of the published literature report experimental results based on objective speech quality metrics such as Perceptual Evaluation of Speech Quality (PESQ)[1], Perceptual Objective Listening Quality Analysis (POLQA) [11], Virtual Speech Quality Objective Listener (ViSQOL) [12], Speech to Distortion Ratio (SDR). These metrics are shown to not correlate well with subjective tests [13]. Few papers do report subjective lab test results, but they are either not statistically significant or the test set is very small.

Common practice in deep learning is to split the dataset into training, validation and test sets. For a SE task, the training set is composed of noisy and clean speech pairs. Noisy speech is usually synthesized by mixing clean speech and noise. Testing the developed models on the synthetic test set gives a heuristic on model performance, but it is not enough to ensure good performance when deployed in real-world conditions. The developed models should be tested on representative real recordings of noisy speech from diverse noisy and reverberant conditions in which speech and noise are captured at the same microphone in similar acoustic conditions. It is hard to simulate these conditions using synthetic data as clean speech and noise signals are captured independently. This makes it difficult for researchers to compare published SE methods and pick the best ones as there is no common test set that is extensive and representative of real-world noisy conditions. Also, there is no reliable subjective test framework that everyone in the research community could use. In [13], we open sourced the Microsoft Scalable Noisy Speech Dataset (MS-SNSD)[2] and an ITU-T P.800 subjective evaluation framework. MS-SNSD includes clean speech and noise recordings and scripts to synthesize noisy speech with augmentation for generating the training set. In addition, a disjoint test set is provided for evaluation. But the test set was missing real recordings and not enough noisy conditions with reverberation. In addition, the P.800 implementation in is missing some crowdsourcing features in P.808 such as hearing and environmental tests, and trapping questions.

The Deep Noise Suppression (DNS) challenge is designed to unify the research work in SE domain by open sourcing the train/test datasets and subjective evaluation framework. We provide large clean speech and noise datasets that are 30 times bigger than MS-SNSD [13]. These datasets are accompanied with configurable scripts to synthesize the training sets. Participants can use any datasets of their choice for training. Half of the test set will be released for the researchers to use during development. The other half will be used as a test set to decide the final competition winners. The online subjective evaluation framework using ITU-T P.808 [14] will be used to compare the submitted SE methods. We also provide the model and inference script for one of the state-of-the-art SE methods as a base algorithm for comparison. This challenge has two tracks based on the computational complexity of the SE method. One track focuses on real-time SE methods and the other track is for non-real-time methods.

---

[1] https://www.itu.int/rec/T-REC-P.862-200102-I/en
[2] https://github.com/microsoft/MS-SNSD

Section 2 describes the datasets. Section 3 describes the base SE method. The online subjective evaluation framework is discussed in section 4. The rules of the challenge and other logistics are described in section 5.

## 2. Datasets

The goal of releasing the clean speech and noise datasets is to provide researchers with the extensive and representative datasets to train their SE models. Previously, we released MS-SNSD [13] with a focus on extensibility. In the recent years, the amount of audio data available over the internet has exploded due to increased content creation on YouTube, smart devices and audiobooks. Though most of these datasets are useful for tasks such as training audio event detectors, automatic speech recognition (ASR) systems etc., most of the SE models need a clean reference, which is not always available. Hence, we synthesize noisy-clean speech pairs.

### 2.1. Clean Speech

The clean speech dataset is derived from the public audio books dataset called Librivox[1]. Librivox corpus is available under the permissive creative commons 4.0 license [15]. Librivox has recordings of volunteers reading over 10,000 public domain audio books in various languages, with majority of which are in English. In total, there are 11,350 speakers. A section of these recordings is of excellent quality, meaning that the speech was recorded using good quality microphones in a silent and less reverberant environments. But there are many audio recordings that are of poor speech quality with speech distortion, background noise and reverberation. Hence, it is important to filter the data based on speech quality.

We used the online subjective test framework ITU-T P.808 [14] to sort the book chapters by subjective quality. The audio chapters in Librivox are of variable length ranging from few seconds to several minutes. We sampled 10 random clips from each book chapter, each 10 seconds in duration. For each clip we had 3 ratings, and the Mean Opinion Score (MOS) across the all clips was used as the book chapter MOS. Figure 1 shows the results, which show the quality spanned from very poor to excellent quality.

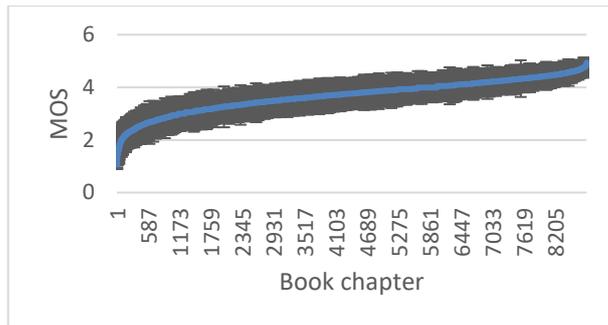

Figure 1: Sorted Librivox P.808 MOS quality with 95% confidence intervals

The upper quartile with respect to MOS was chosen as our clean speech dataset, which are top 25% of the clips with MOS as a metric. The upper quartile comprised of audio chapters with $4.3 \leq MOS \leq 5$. We removed clips from speakers with less than 15 minutes of speech. The resulting dataset has 500 hours of speech from 2150 speakers. All the filtered clips are then split into segments of 10 seconds. In total, we use approx. 441 hours of clean speech for generating the upper quartile subset.

### 2.2. Noise Dataset

The noise clips were selected from Audioset[2] [16] and Freesound[3]. Audioset is a collection of about 2 million human-labeled 10s sound clips drawn from YouTube videos and belong to about 600 audio events. Like the Librivox data, certain audio event classes are overrepresented. For example, there are over a million clips with audio classes music and speech and less than 200 clips for classes such as toothbrush, creak etc. Approximately, 42% of the clips have single class, but the rest may have 2 to 15 labels. Hence, we developed a sampling approach to balance the dataset in such a way that each class has at least 500 clips. We also used a speech activity detector to remove the clips with any kind of speech activity. The reason is to avoid suppression of speech by the noise suppression model trained to suppress speech like noise. The resulting dataset has about 150 audio classes and 60,000 clips. We also augmented an additional 10,000 noise clips downloaded from Freesound and DEMAND databases [17]. The chosen noise types are more relevant to VOIP applications.

### 2.3. Noisy Speech

The clean speech and noise datasets can be found in the repo[4]. The noisy speech database is created by adding clean speech and noise at various Signal to Noise Ratio (SNR) levels. We compute segmental SNR using segments in which both speech and noise are active. This is to avoid overshooting of amplitude levels in impulsive noise types such as door shutting, clatter, dog barking etc. We synthesize 30s long clips by augmenting clean speech utterances and noise. The SNR levels are sampled from a uniform distribution between 0 and

---

[1] https://librivox.org/
[2] https://research.google.com/audioset/
[3] https://freesound.org/
[4] https://github.com/microsoft/DNS-Challenge/tree/master/datasets

40 dB. The mixed signal is then set to target Root Mean Square (RMS) level sampled from a uniform distribution between -15 dBFS and -35 dBFS. The data generation scripts are open sourced in the DNS-Challenge repo[1].

### 2.4. Test clips

We are open sourcing a new test set that comprises both synthetic and real recordings. It is a general practice to evaluate the SE method on a synthetic test set. But a synthetic test set is not a good representative of what we observe in the wild. The synthetic test set might be useful in tuning the model during development phase using objective metrics such as PESQ and POLQA that require clean reference. Generally, in synthetic data the original clean speech and noise are collected in different acoustic conditions using two different microphones and are mixed to form noisy speech. With real recordings the clean speech and noise are captured at the same microphone and acoustic conditions.

The test set is divided into 4 categories with 300 clips in each:
1. Synthetic clips without reverb
2. Synthetic clips with reverb
3. Real recordings collected internally at Microsoft
4. Real recordings from Audioset

For synthetic test clips, we used Graz University's clean speech dataset [18] which consists of 4,270 recorded sentences spoken by 20 speakers. For the synthetic clips with reverb, we add reverberation to the clean files using the room impulse responses recorded internally at Microsoft with RT60 ranging from 300ms to 1300ms. We sampled 15 clips from 12 noise categories we deem highly important for VoIP scenarios to synthesize 180 noisy clips. The 12 categories are fan, air conditioner, typing, door shutting, clatter noise, car, munching, creaking chair, breathing, copy machine, baby crying and barking. The remaining 120 noise clips were randomly sampled from the remaining 100+ noise classes. The SNR levels were sampled from a uniform distribution between 0 dB and 25 dB. The real recordings collected internally at Microsoft consist of recorded noisy speech in various open office and conference rooms noisy conditions. We hand-picked 300 audio clips with speech mixed in noise from AudioSet that we felt are relevant to audio calls we experience in noisy conditions.

## 3. Baseline SE method

As a baseline, we will use the recently developed SE method from **Error! Reference source not found.**, which is based on Recurrent Neural Network (RNN). For ease of reference, we will call this method as Noise Suppression Net (NSNet). This method uses log power spectra as input to predict the enhancement gain per frame using a learning machine based on Gated Recurrent Units (GRU) and fully connected layers. Please refer to the paper for more details of the method.

NSNet is computationally efficient. It only takes 0.16ms to enhance a 20ms frame on an Intel quad core i5 machine using the ONNX run time v1.1[2]. It is subjectively evaluated using a large test set showing improvement over conventional SE method.

We have open sourced the inference script and the model in ONNX format in the challenge DNS-Challenge repo[3].

## 4. Online Subjective Evaluation Framework ITU-T P.808

We use the ITU-T P.808 Subjective Evaluation of Speech Quality with a Crowdsourcing Approach [14] methodology to evaluate and compare SE methods using Absolute Category Ratings (ACR) to estimate a Mean Opinion Score (MOS). We created an open source[4] implementation of P.808 using the Amazon Mechanical Turk platform. This system has the following features/attributes:

- Raters are first qualified using a hearing and environmental test before they can start rating clips. This ensures raters have a sufficient hearing ability, a good quality listening device, and a quiet environment to do ratings in. Our implementation allows raters to start rating clips immediately after being qualified which increased the rating speed by ~5X compared to having a separate qualification stage.
- Raters are given several training examples but are not screened using the results; the training is used for anchoring purposes.
- Audio clips are rated in groups of clips (e.g., N=10). Each group includes a gold clip with known ground truth (e.g. a clean or very poor clip) and a trapping question (e.g., "This is an interruption: Please select option 2"). The gold and trapping questions are used for filtering out "spam" raters who are not paying attention.
- Every hour raters are also given comparison rating test using gold samples (e.g., which is better, A or B) to verify their environment is still valid to do ratings in.
- Raters are restricted to rating a limited number of clips per P.808 recommendations to reduce rater fatigue.

To validate the measurement system accuracy, we rated the ITU Supplement 23 Experiment 3 [20] dataset which has published lab-based MOS results. The system gives a Spearman correlation coefficient of 0.93 to the lab results given in ITU Supplement 23 (MOS is computed per test condition). To validate the system repeatability, we ran the ITU Supplement 23 twice (on separate days, with <10% overlapped raters, and 1/10th the ratings as Run 1) and the results were similar (see Table 1).

|  | $\rho$ |
| --- | --- |
| ITU Supplement 23 Run 1 | 0.93 |
| ITU Supplement 23 Run 2 | 0.87 |

Table 1: P.808 Spearman rank correlation with ITU Supplement 23 Experiment 3

---

[1] https://github.com/microsoft/DNS-Challenge
[2] https://github.com/microsoft/onnxruntime
[3] https://github.com/microsoft/DNS-Challenge/tree/master/NSNet-baseline
[4] https://github.com/microsoft/P.808

# 5. DNS Challenge Rules and Schedule

## 5.1. Rules

All participants should adhere to the following rules to be eligible for the challenge.
1. Participants can use any training datasets of their choice. They can also augment additional data to the provided dataset. They can mix clean speech and noise in any way that improves the performance of their SE method. We also encourage participants to open source their datasets so that it helps the greater research community.
2. Participants can test their developed methods on any test set during the development phase. But we encourage them to use our test set as it is extensive and is a good representation of real-world scenarios.
3. Every participating SE method will fall in one of the two tracks depending on the computational complexity. Track 1 is focused on low computational complexity. The algorithm should take less than $T/2$ (in ms) to process a frame of size $T$ (in ms) on an Intel Core i5 quad core machine clocked at 2.4 GHz or equivalent processors. Frame length $T$ should be less than or equal to 40ms. Track 2 does not have any constraints on computational time so that researchers can explore deeper models to attain exceptional speech quality.
4. In both the tracks, the SE method can have a maximum of 40ms look ahead. To infer the current frame $T$ (in ms), the algorithm can access any number of past frames but only 40ms of future frames ($T$+40ms).
5. Winners will be picked from each track based on the subjective speech quality evaluated on the blind test set using ITU-T P.808 framework.
6. The blind test set will be made available to the participants on March 18[th]. Participants should send the enhanced clips using their developed models to the organizers. We will use the submitted clips with no alteration to conduct ITU-T P.808 subjective evaluation and pick the winners based on the results. Participants are forbidden from using the blind test set to retrain or tweak their models. They should not submit enhanced clips using other noise suppression methods that they are not submitting to INTERSPEECH 2020. Failing to adhere to these rules will lead to disqualification from the challenge.
7. Participants should report the computational complexity of their model in terms of the number of parameters and the time it takes to infer a frame on a particular CPU (preferably Intel Core i5 quad core machine clocked at 2.4 GHz). Among the submitted proposals differing by less than 0.1 MOS, the lower complexity model will be given higher ranking.
8. Each participating team is expected to submit an INTERSPEECH paper that summarizes the research efforts and provide all the details to ensure reproducibility. Authors may choose to report additional objective/subjective metrics in their paper.
9. Submitted papers will undergo standard peer-review process of INTERSPEECH 2020. The paper needs to be accepted to the conference for the participants to be eligible for the challenge.
10. Relevant papers submitted to regular session might be included in this challenge for fostering in-depth discussions.

## 5.2. Timeline

- **January 20[th], 2020:** Release of the datasets and scripts for training and testing.
- **March 30[th], 2020:** Complimentary P.808 Subjective evaluation.
- **April 22[nd], 2020**: Release of the blind test set.
- **April 27[th], 2020**: Deadline for participants to submit their enhanced clips for P.808 evaluation on the blind test set.
- **May 4[th], 2020:** Organizers will notify the participants about the results.
- **May 8[th], 2020:** Regular paper submission deadline for Interspeech 2020.
- **July 24[th], 2020:** Paper acceptance/rejection notification
- **July 31[st], 2020:** Notification of the winners with winner instructions, including a prize claim deadline.

## 5.3. Support

Participating teams may email organizers at **dns_challenge@microsoft.com** questions or need any clarification about any aspect of the challenge.

# 6. Conclusions

This challenge aims to promote real-time single microphone noise suppression for exceptional subjective speech quality. We are providing training and test datasets for researchers to train their models. The final evaluation will be done using ITU-T P.808

# 7. Acknowledgements

The P.808 implementation was written by Babak Naderi.

# 8. References

[1]	Y. Ephraim and D. Malah, "Speech enhancement using a minimum mean-square error log-spectral amplitude estimator," *IEEE Trans. Acoust. Speech Signal Process.*, vol. 33, no. 2, pp. 443–445, Apr. 1985, doi: 10.1109/TASSP.1985.1164550.
[2]	C. Karadagur Ananda Reddy, N. Shankar, G. Shreedhar Bhat, R. Charan, and I. Panahi, "An Individualized Super-Gaussian Single Microphone Speech Enhancement for Hearing Aid Users With Smartphone as an Assistive Device," *IEEE Signal Process. Lett.*, vol. 24, no. 11, pp. 1601–1605, Nov. 2017, doi: 10.1109/LSP.2017.2750979.
[3]	P. J. Wolfe and S. J. Godsill, "Simple alternatives to the Ephraim and Malah suppression rule for speech enhancement," in *Proceedings of the 11th IEEE Signal Processing Workshop on Statistical Signal Processing (Cat. No.01TH8563)*, Singapore, 2001, pp. 496–499, doi: 10.1109/SSP.2001.955331.
[4]	T. Lotter and P. Vary, "Speech Enhancement by MAP Spectral Amplitude Estimation Using a Super-Gaussian Speech Model," *EURASIP J. Adv. Signal Process.*, vol. 2005, no. 7, p. 354850, Dec. 2005, doi: 10.1155/ASP.2005.1110.
[5]	S. Srinivasan, J. Samuelsson, and W. B. Kleijn, "Codebook-Based Bayesian Speech Enhancement for Nonstationary Environments," *IEEE Trans. Audio Speech Lang. Process.*, vol. 15, no. 2, pp. 441–452, Feb. 2007, doi: 10.1109/TASL.2006.881696.
[6]	Y. Xu, J. Du, L.-R. Dai, and C.-H. Lee, "A Regression Approach to Speech Enhancement Based on Deep Neural Networks," *IEEEACM Trans. Audio Speech Lang. Process.*, vol. 23, no. 1, pp. 7–19, Jan. 2015, doi: 10.1109/TASLP.2014.2364452.

[7] Y. Luo and N. Mesgarani, "Conv-TasNet: Surpassing Ideal Time-Frequency Magnitude Masking for Speech Separation," *IEEEACM Trans. Audio Speech Lang. Process.*, vol. 27, no. 8, pp. 1256–1266, Aug. 2019, doi: 10.1109/TASLP.2019.2915167.

[8] A. Pandey and D. Wang, "TCNN: Temporal Convolutional Neural Network for Real-time Speech Enhancement in the Time Domain," in *ICASSP 2019 - 2019 IEEE International Conference on Acoustics, Speech and Signal Processing (ICASSP)*, Brighton, United Kingdom, May 2019, pp. 6875–6879, doi: 10.1109/ICASSP.2019.8683634.

[9] D. Yin, C. Luo, Z. Xiong, and W. Zeng, "PHASEN: A Phase-and-Harmonics-Aware Speech Enhancement Network," *ArXiv191104697 Cs Eess*, Nov. 2019, Accessed: Jan. 06, 2020. [Online]. Available: http://arxiv.org/abs/1911.04697.

[10] A. Ephrat *et al.*, "Looking to Listen at the Cocktail Party: A Speaker-Independent Audio-Visual Model for Speech Separation," *ACM Trans. Graph.*, vol. 37, no. 4, pp. 1–11, Jul. 2018, doi: 10.1145/3197517.3201357.

[11] J. G. Beerends, M. Obermann, R. Ullmann, J. Pomy, and M. Keyhl, "Perceptual Objective Listening Quality Assessment (POLQA), The Third Generation ITU-T Standard for End-to-End Speech Quality Measurement Part I–Temporal Alignment," *J Audio Eng Soc*, vol. 61, no. 6, p. 19, 2013.

[12] A. Hines, J. Skoglund, A. C. Kokaram, and N. Harte, "ViSQOL: an objective speech quality model," *Eurasip J. Audio Speech Music Process.*, no. 1, p. 13, 2015, doi: 10.1186/s13636-015-0054-9.

[13] C. K. A. Reddy, E. Beyrami, J. Pool, R. Cutler, S. Srinivasan, and J. Gehrke, "A Scalable Noisy Speech Dataset and Online Subjective Test Framework," in *Interspeech 2019*, Sep. 2019, pp. 1816–1820, doi: 10.21437/Interspeech.2019-3087.

[14] "ITU-T P.808: Subjective evaluation of speech quality with a crowdsourcing approach," 2018.

[15] V. Panayotov, G. Chen, D. Povey, and S. Khudanpur, "Librispeech: An ASR corpus based on public domain audio books," in *2015 IEEE International Conference on Acoustics, Speech and Signal Processing (ICASSP)*, Apr. 2015, pp. 5206–5210, doi: 10.1109/ICASSP.2015.7178964.

[16] J. F. Gemmeke *et al.*, "Audio Set: An ontology and human-labeled dataset for audio events," in *2017 IEEE International Conference on Acoustics, Speech and Signal Processing (ICASSP)*, New Orleans, LA, Mar. 2017, pp. 776–780, doi: 10.1109/ICASSP.2017.7952261.

[17] J. Thiemann, N. Ito, and E. Vincent, "The Diverse Environments Multi-channel Acoustic Noise Database (DEMAND): A database of multichannel environmental noise recordings," presented at the ICA 2013 Montreal, Montreal, Canada, 2013, pp. 035081–035081, doi: 10.1121/1.4799597.

[18] G. Pirker, M. Wohlmayr, S. Petrik, and F. Pernkopf, "A Pitch Tracking Corpus with Evaluation on Multipitch Tracking Scenario," p. 4.

[19] Y. Xia, S. Braun, C. K. A. Reddy, H. Dubey, R. Cutler, and I. Tashev, "Weighted Speech Distortion Losses for Neural-Network-Based Real-Time Speech Enhancement," in *ICASSP 2020 - 2020 IEEE International Conference on Acoustics, Speech and Signal Processing (ICASSP)*, Barcelona, Spain, May 2020, pp. 871–875, doi: 10.1109/ICASSP40776.2020.9054254.

[20] "ITU-T Supplement 23 ITU-T coded-speech database Supplement 23 to ITU-T P-series Recommendations (Previously CCITT Recommendations)," 1998.